\documentclass[12pt]{article}
\usepackage{amsmath}

\newcommand{\be}{\begin{equation}}
\newcommand{\ee}{\end{equation}}
\usepackage{graphicx}

\def\eqn#1{$$#1$$}

\def\eqnn#1#2{
	\begin{equation}#2\label{#1}
	\end{equation}}

\def\elnn#1#2{\begin{align}\begin{split}#2\label{#1}
	\end{split}\end{align}}
\def\elnnn#1{\begin{align}#1\end{align}}

\begin{document}
\baselineskip=24 pt

\begin{center}

{\large {\bf General Relativity Resolves Galactic Rotation Without Exotic 
Dark Matter}}

\end{center}

\vskip1.5truecm 

\begin{center}
  F. I. Cooperstock and S. Tieu \\
{\small \it Department of Physics and Astronomy, University 
of Victoria} \\
{\small \it P.O. Box 3055, Victoria, B.C. V8W 3P6 (Canada)}\\

{\small \it e-mail addresses: cooperstock@phys.uvic.ca, stieu@uvic.ca}
\end{center} 

\begin{abstract} 
A galaxy is modeled as a stationary axially symmetric pressure-free 
fluid in general relativity. For the weak gravitational fields under 
consideration, the field equations and the equations of motion 
ultimately lead to one linear and one nonlinear equation relating the 
angular velocity to the fluid density. It is shown that the rotation 
curves for the Milky Way, NGC 3031, NGC 3198 and NGC 7331 are consistent with 
the mass density distributions of the visible matter concentrated in 
flattened disks.
Thus the need for a massive halo of exotic dark matter is removed.
For these galaxies we determine the mass density for the luminous threshold as $10^{-21.75}$ kg$\cdot$m$^{-3}$.

\end{abstract} 
\vspace*{1truecm}

\begin{center}
\textit{Subject headings}: galaxies: kinematics and 
dynamics-gravitation-relativity-dark matter\\
\mbox{ }\\
\end{center}

\section{Introduction}
The problem of accounting for the observed essentially flat galactic 
rotation curves has been a central issue in astrophysics. There has 
been much speculation over the question of the nature of the dark 
matter that is believed to be required for the consistency of the 
observations with Newtonian gravitational theory. Clearly the issue 
is of paramount importance given that the dark matter is said to 
constitute the dominant constituent of a galactic mass \cite{BT}. The 
dark matter enigma has served as a spur for particle theorists to 
devise acceptable candidates for its constitution. While physicists 
and astrophysicists have pondered over the issue, other researchers 
have devised new theories of gravity to account for the observations 
(see for example \cite{bek}). However the latter approaches, 
imaginative as they may be, have met with understandable skepticism, 
having been devised solely for the purpose of the task at hand. 
General relativity remains the preferred theory of gravity with 
Newtonian theory as its limit. General relativity has been successful 
in every test that it has encountered, going beyond Newtonian theory 
where required.

It is understandable that the conventional gravity approach has 
focused upon Newtonian theory in the study of galactic dynamics as 
the galactic field is weak (apart from the deep core regions where 
black holes are said to reside) and the motions are non-relativistic 
($v \ll{c}$). It was this approach that led to the inconsistency 
between the theoretical Newtonian-based predictions and the 
observations of the visible sources. To reconcile the theory with the 
observations, researchers subsequently concluded that dark matter 
must be present around galaxies in vast massive halos that constitute 
the great bulk of the galactic masses \footnote{
	See however \cite{keet} who  argues for a much less massive
	halo based upon gravitational lensing  data.}.
However, in dismissing general relativity in favor of 
Newtonian gravitational theory for the study of galactic dynamics, 
insufficient attention has been paid to the fact that the stars that 
compose the galaxies are essentially in motion under gravity alone 
(``gravitationally bound''). It has been known since the time of 
Eddington that the gravitationally bound problem in general 
relativity is an intrinsically non-linear problem even when the 
conditions are such that the field is weak and the motions are 
non-relativistic, at least in the time-dependent case. \textit{Most 
significantly, we have found that under these conditions, the general 
relativistic analysis of the problem is also non-linear for the 
stationary (non-time-dependent) case at hand.} Thus the intrinsically 
linear Newtonian-based approach used to this point has been 
inadequate for the description of the galactic dynamics and 
Einstein's general relativity should be brought into the analysis 
within the framework of established gravitational theory\footnote{
	Actually within the framework of Newtonian theory, it is 
	possible to define an ``effective'' potential (see for 
	example \cite{BT} page 136) to incorporate the centrifugal 
	acceleration in a rotating coordinate system with a given 
	angular velocity. Since this contains the square of the 
	angular velocity of the rotating frame, there is already 
	the hint of non-linearity present.  However, in what 
	follows in general relativity, we will see the 
	non-linearity related to the angular velocity as a 
	\textit{variable} function. Moreover, for a system in 
	rotation, this non-linearity cannot be removed globally.
}.  This is an essential departure from conventional thinking on the 
subject and it leads to major consequences as we discuss in what 
follows. We will demonstrate that via general relativity, the 
generating potentials producing the observed flattened galactic 
rotation curves are necessarily linked to the mass density
distributions of the flattened disks, obviating any necessity for
dark matter halos in the total galactic composition. We will also present the indicator that the threshold for luminosity occurs at a density of $10^{-21.75}$ kg$\cdot$m$^{-3}$.

\section{The Model Galaxy}

Within the context of Newtonian theory, Mestel \cite{mes} considered 
a special rotating disk with surface density inversely proportional 
to radius. Using a disk potential with Bessel functions that we will 
also use in what follows but in quite a different manner, he found 
that it leads to an absolutely flat galactic rotation velocity
curve.\footnote{
	This is also the case for the MOND \cite{bek} model.
} 
Interestingly, the gradient of the potential in this, as in all 
Newtonian treatments, relates to acceleration whereas in the general 
relativistic treatment, we will show that the gradient of a 
``generating potential'' gives the tangential velocity (\ref{Eq11}).  

To model a galaxy in its simplest form in terms of its essential 
characteristics, we consider a uniformly rotating fluid without 
pressure and symmetric about its axis of rotation. We do so within 
the context of general relativity. The stationary axially symmetric 
metric can be described in generality in the form
\eqnn{Eq1}{
	ds^2 =	-e^{\nu-w}( udz^2+dr^2)
		-r^2 e^{-w} d\phi^2+e^w(cdt-Nd\phi)^2
}
where $u$, $\nu$, $w$ and $N$ are functions of cylindrical polar 
coordinates $r$, $z$.  It is easy to show that to the order required, 
$u$ can be taken to be unity.  It is most simple to work in the frame 
that is co-moving with the matter
\eqnn{Eq2}{
	U^i = {\delta}_0^i
}
where $U^i$ is the 4-velocity\footnote{
	This is reminiscent of the standard approach that is 
	followed for FRW cosmologies. However, the FRW spacetimes 
	are homogeneous and they are not stationary
}.
This was done in the pioneering paper by van Stockum  \cite{vs} who 
set $w=0$ from the outset\footnote{
	Interestingly, the geodesic equations imply that 
	$w=constant$ (which can be taken to be zero as in 
	\cite{vs}) even for the \textit{exact} Einstein field 
	equations as studied in \cite{vs} (see (\ref{Eq6}) and 
	(\ref{Eq7})).
}.  As in \cite{Bonnor}, we perform a purely \textit{local}
($r$, $z$ held fixed) transformation
\eqnn{Eq3}{
	\bar{\phi} = \phi + \omega(r,z)\,t
}
that locally diagonalizes the metric. In this manner, we are able to 
deduce the local angular velocity $\omega$ and the tangential 
velocity $V$ as
\elnn{Eq4}{
	\omega
	&= \frac{Nce^w}{r^2e^{-w}-N^2e^w}
	   \approx  \frac{Nc}{r^2},  \\
	V &=\omega r
}
with the approximate value applicable for the weak fields under 
consideration.  The Einstein field equations to order $G$ are\footnote{
	This is a loose notation favored by many relativists but adequate
	for our purposes here as a smallness parameter.
}
\elnn{Eq5}{
	2r\nu_r+ N_r^2-N_z^2 &=0, \\
	r\nu_z +N_r N_z &=0,  \\
	N_r^2 + N_z^2 +2r^2(\nu_{rr}+\nu_{zz}) &=0, \\
	N_{rr} +N_{zz} - \frac{N_r}{r}&=0, \\
	\left(w_{rr} +w_{zz} +\frac{w_r}{r}\right)
	+ \frac{3}{4}r^{-2} (N_r^2 + N_z^2)& \\ 
 	+ Nr^{-2}\left(N_{rr} +N_{zz} -\frac{N_r}{r}\right)
	- \frac{1}{2}(\nu_{rr}+\nu_{zz})
	&= \frac{8{\pi}G\rho}{c^2}
}
where $G$ is the gravitational constant and $\rho$ is the mass 
density.  Subscripts denote partial differentiation with respect to 
the indicated variable.  These equations are easily combined to yield
\eqnn{Eq5a}{
	\nabla^2 w +\frac{N_r^2+N_z^2}{r^2}=\frac{8\pi G\rho}{c^2}
}
where
\eqnn{Eq5b}{
	\nabla^2 w \equiv w_{rr} + w_{zz} + \frac{w_r}{r}
}
and $\nu$ would be determined by quadratures.
Since the pressure-free fluid elements must satisfy the geodesic 
equation as their equation of motion
\eqnn{Eq6}{
	\frac{dU^i}{ds} +\Gamma^i_{kl}U^k U^l=0
}
we find using (\ref{Eq1}) and (\ref{Eq2}) that
\eqnn{Eq7}{
	w_r=w_z=0
}
and hence 
\eqnn{Eq8}{
	\nabla^2 w=0
}
within the fluid\footnote{
	Normally, the fall-off of $w$ with
	$R \equiv \sqrt{r^2 +z^2}$ is used to derive the total mass of
	an isolated system. However, $w$ is constant in this system
	of coordinates by (\ref{Eq7}) and we cannot do so here.
	The $w$ constancy  does not imply 
	that that the mass is zero. In other (non-co-moving) 
	coordinate systems, $w$ would be seen to be variable. With 
	the field being weak and the system being non-relativistic, 
	the mass is well-approximated simply by the integral of 
	$\rho$ over \textit{coordinate} volume. Moreover, we will 
	choose solutions that are free of singularities and hence 
	free of the ambiguities present in \cite {Bonnor}.
}.  It is to be noted that it is the \textit{freely gravitating 
motion} of the source material (the stars) in conjunction with the 
choice of co-moving coordinates (\ref{Eq2}) that leads to the 
constancy of $w$ within the source. Had there been pressure, $w$ 
would have been variable\footnote{
	Even in freely gravitating motion, $w$ would have been 
	variable had we opted for non-co-moving coordinates.
}.  With this freely gravitating constraint, the interior field
equations for $N$ and $\rho$ are reduced to
\elnnn{
	N_{rr} + N_{zz} - \frac{N_r}{r} &=0
	\label{Eq9a} \\
	\frac{N_r^2 + N_z^2}{r^2} &= \frac{8{\pi}G\rho}{c^2}
	\label{Eq9b}
}
Note that from both the field equation for $\rho$ and the expression 
for $\omega$ that $N$ is of order $G^{1/2}$.  The non-linearity of 
the galactic dynamical problem is manifest through the
non-linear relation\footnote{
	While we have eliminated $w$ using (\ref{Eq6}) to get 
	(\ref{Eq9b}) by choice of co-moving coordinates, $N$ cannot 
	be eliminated and hence non-linearity is intrinsic to the 
	study of the galactic dynamics.
}
between the functions $\rho$ and $N$.  
Rotation under freely gravitating motion is the key here. By 
contrast, for time-independence in the non-rotating problem, there 
must be pressure present to maintain a static configuration, $N$ 
vanishes for vanishing $\omega$ and $\nabla^2 w$ is non-zero yielding 
the familiar Poisson equation of Newtonian gravity. In the present 
case, it is the \textit{rotation} via the function $N$ that connects 
directly to the density and the now non-linear equation is in sharp 
contrast to the linear Poisson equation.  Interestingly, (\ref{Eq9a}) 
can be expressed as 
\eqnn{Eq10}{
	\nabla^2\Phi =0
}
where
\eqnn{Eq10a}{
	\Phi \equiv \int\frac{N}{r}dr
}
and hence flat-space harmonic functions $\Phi$ are the generators of the 
axially symmetric stationary pressure-free weak fields that we
seek\footnote{
	In fact Winicour \cite{Winicour} has shown that all such sources,
	even when the fields are strong, are generated by such flat-space
	harmonic functions.
}.  Using (\ref{Eq4}) and (\ref{Eq10a}), we have the expression for the 
tangential velocity of the distribution
\elnn{Eq11}{
	V&=c\frac{N}{r} \\
	&=c\frac{\partial {\Phi}}{\partial{r}}
}

\section{Modeling the Observed Galactic Rotation Curves}

\begin{figure}
\begin{center}
\includegraphics[width=3in]{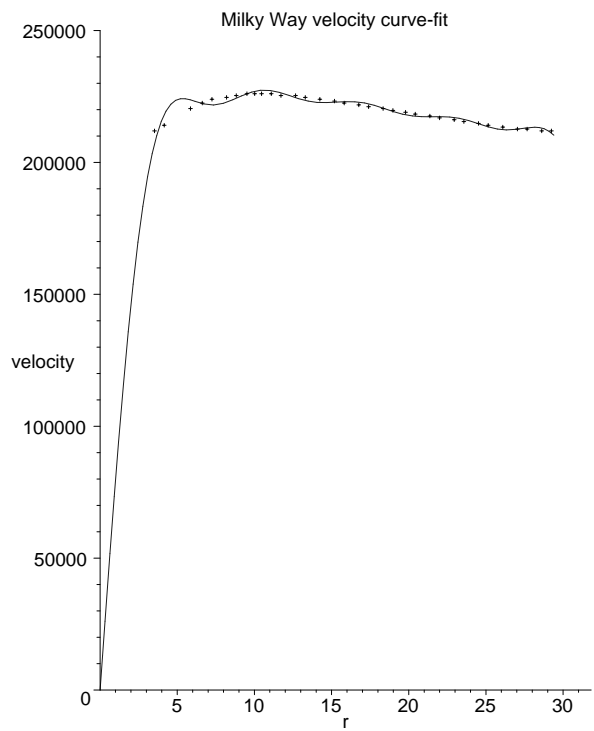}
\end{center}
\caption{
	Velocity curve-fit for the Milky Way
	in units of m/s vs Kpc.
}
\label{fig:milkywayvelocity}
\end{figure}

\begin{figure}
\begin{center}
\includegraphics[width=3in]{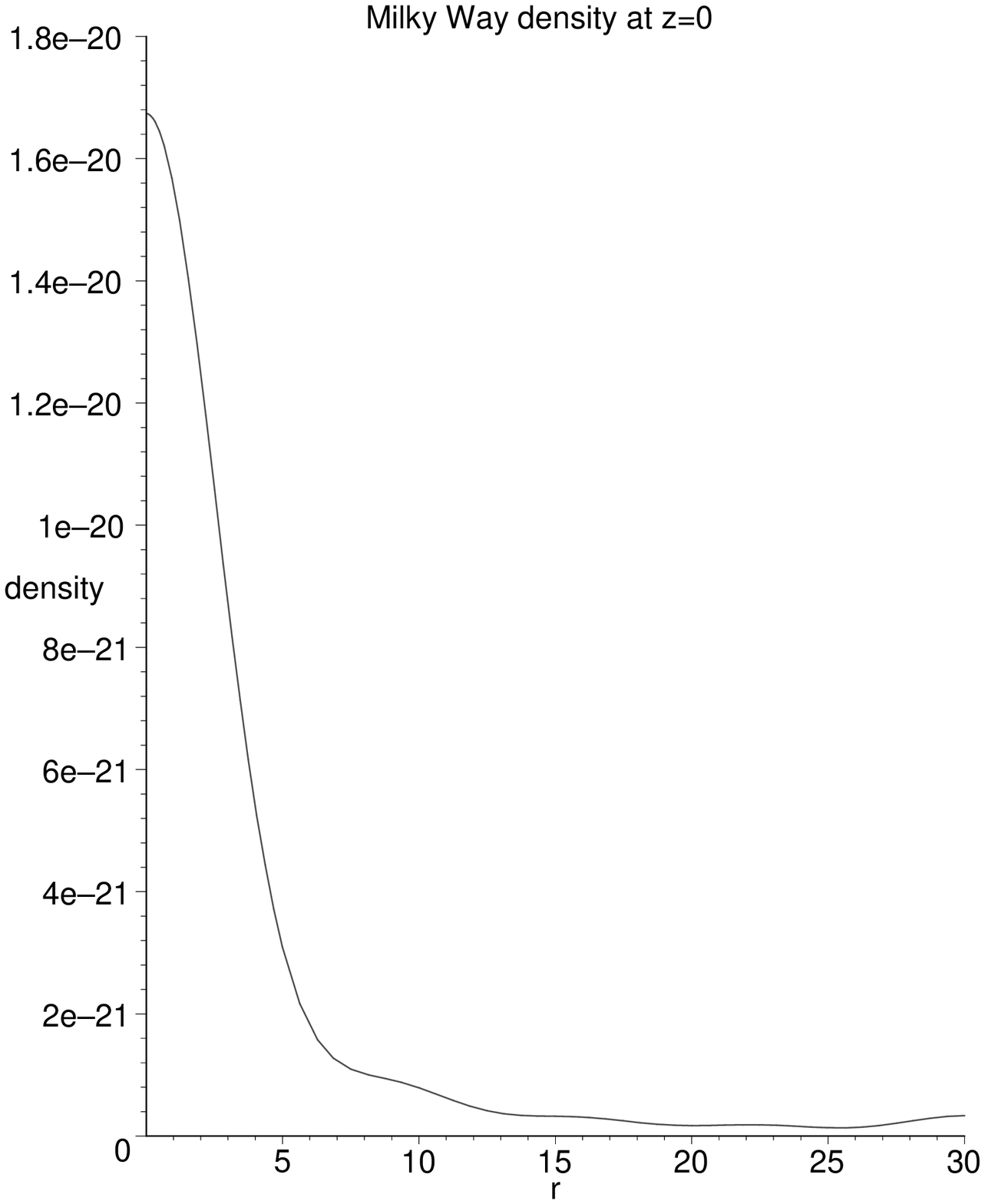}
\includegraphics[width=3in]{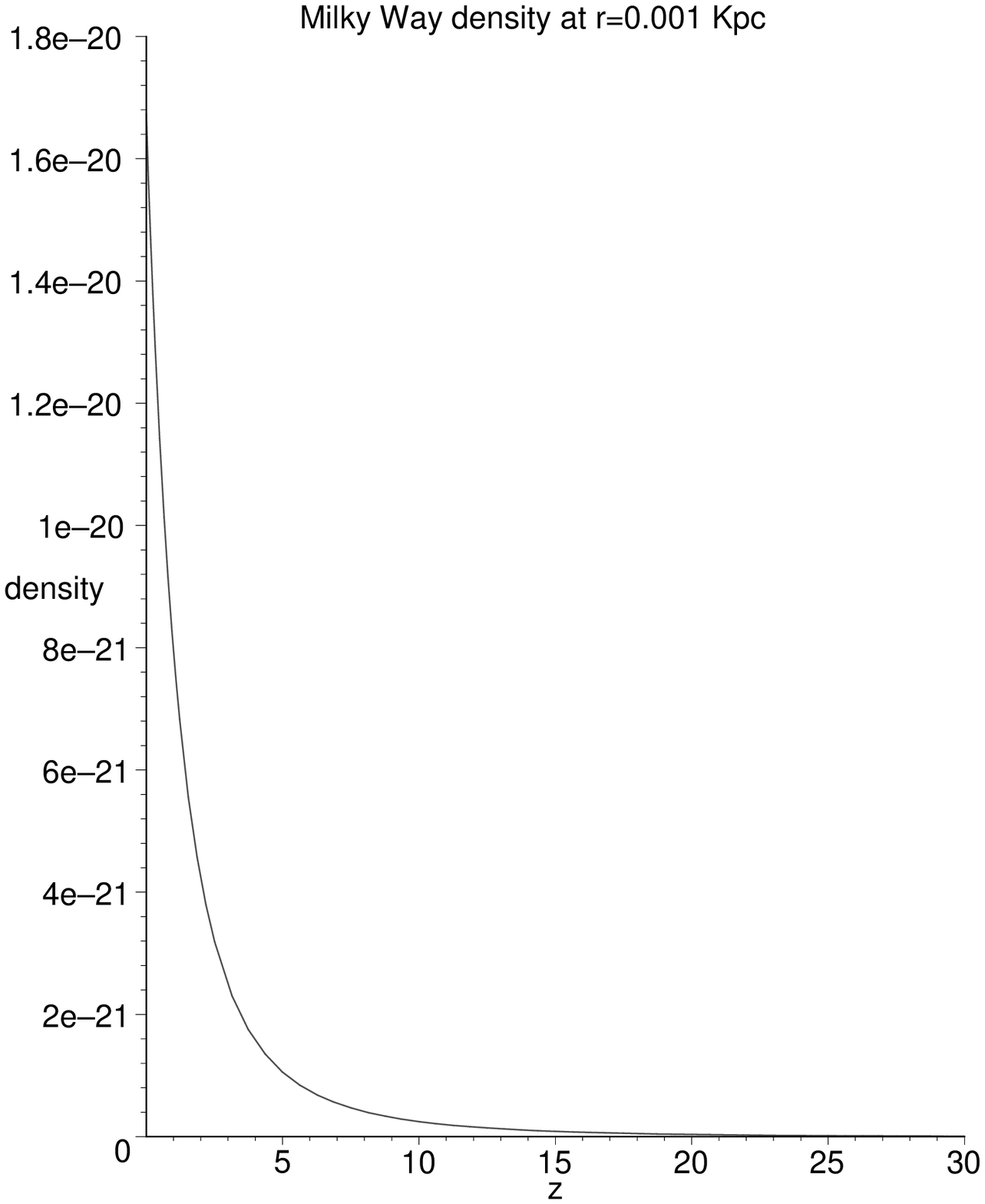}
\end{center}
\caption{
	Derived density profiles in units of kg/m$^3$
	for the Milky Way at $z=0$ (left)
	and $r=0.001$ Kpc (right).
}
\label{fig:milkywaydensity}
\end{figure}

Since the field equation for $\rho$ is non-linear, the simpler way to 
proceed in galactic modeling is to first find the required generating 
potential $\Phi$ and from this, derive an appropriate function $N$ 
for the galaxy that is being analyzed. With $N$ found, (\ref{Eq9b}) 
yields the density distribution. If this is in accord with 
observations, the efficacy of the approach is established. Every 
galaxy is different and each requires its own composing elements to 
build the generating potential. In cylindrical polar coordinates, 
separation of variables yields the following solution for $\Phi$
in (\ref{Eq10}):
\eqnn{Eq12}{
	\Phi = Ce^{-k\mid z \mid}J_0(kr)
}
where $J_0$ is the Bessel function $n=0$ of Bessel $J_n(kr)$ and $C$
is an arbitrary constant. \footnote{
	The absolute value of $z$ must be used to provide the 
	proper reflection of the distribution for negative $z$. 
	While this produces a discontinuity in $N_z$ at $z=0$, it 
	is important to note that this has no physical consequence 
	since $N_z$ enters as a square in the density and $N_z$ 
	does not play a role in the equations of motion. Moreover, 
	the metric itself is continuous. This is analogous to the 
	Schwarzschild constant density sphere problem that leads to 
	a discontinuity in metric derivative across the 
	matter-vacuum interface in Schwarzschild coordinates. In 
	principle, other coordinates could be found to render the 
	metric and its first partial derivatives globally 
	continuous but this would be counter-productive as it would
	unnecessarily complicate the mathematics. As in FRW, our co-moving coordinates simplify the analysis.
} We use the linearity of (\ref{Eq10}) to express the general solution of 
this form as a linear superposition 
\eqnn{Eq13}{
	\Phi = \sum_{n}C_ne^{-k_n |z|}J_0(k_nr)
}
with $n$ chosen appropriately for the desired level of accuracy.
From (\ref{Eq13}) and (\ref{Eq4}), the tangential velocity \footnote{
	$dJ_0(x)/dx= - J_1(x)$ from \cite{ford}.
} is
\eqnn{Eq14}{
	V= -c\sum_{n} k_n C_n e^{-k_n |z|}J_1(k_nr)
}
With the $k_n$ chosen so that the $J_0(k_nr)$ terms are orthogonal
\footnote{
	Just as the $\sin kx$ functions are orthogonal for integer 
	$k$, the Bessel functions $J_0(kr)$ have their own 
	orthogonality relation: $\int_{0}^{1} J_0(k_nr)J_0(k_mr)rdr 
	\propto \delta_{mn}$ where $k_n$ are the zeros of $J_0$ at 
	the limits of integration. This orthogonality condition is 
	on $\Phi$ rather than on $V$ because the differential 
	equation dictates the integral condition.
}
to each other, we have found that only 10 functions with parameters 
$C_n$, $n\in\{1\dots 10\}$ suffice to provide an excellent fit\footnote{
	It should be noted that unlike typical velocity curve fits 
	that allow arbitrary velocity functions, our curve fits are 
	constrained by the demand that they be created from 
	derivatives of harmonic functions.
}
to the velocity curve for the Milky Way.  The details are provided in 
the Appendix and the curve fit is shown in Figure 
\ref{fig:milkywayvelocity}.
\footnote{
	Note that the $J_1(x)$ Bessel functions are $0$ at $x=0$ 
	and oscillate with decreasing amplitude, falling as 
	$1/\sqrt{x}$ asymptotically \cite{ford}, as desired for 
	merging with Keplerian behavior at infinity. Also, the 
	present curves drop as $r$ approaches $0$.  This is in 
	contrast to the Mestel \cite{mes} and MOND \cite{bek}
	curves that are flat everywhere.
} From (\ref{Eq11}) and (\ref{Eq14}), the $N$ function is determined in 
detail and from (\ref{Eq9b}), the density distribution. This is shown 
in Figure \ref{fig:milkywaydensity} as a function of $r$ at $z=0$ as 
well as a function of $z$ at $r=0.001$ Kpc.  We see that the 
distribution is a flattened disk with good correlation with the 
observed density data for the Milky Way.
The integrated mass is found to be $21 \times 10^{10}M_\odot$ which
is at the lower end of the estimated mass range of
$20 \times 10^{10}M_\odot$ to $60 \times 10^{10}M_\odot$
as established by various researchers. 
It is to be noted that the approximation scheme would break down in 
the region of the galactic core should the core harbor a black hole 
or even a naked singularity (see e.g. \cite{coop}). \textit{Most significantly, our 
correlation of the flat velocity curve is achieved with disk mass of an
order of magnitude smaller than the envisaged halo mass of exotic dark 
matter.} (See e.g. \cite{clew} for proposed values of extended halo masses.)  General relativity does not distinguish between the luminous 
and non-luminous contributions. The $\rho$ density distribution 
deduced is derived from the totality of the two. Any substantial 
amount of non-luminous matter would necessarily lie in the flattened 
region close to $z=0$ because this is the region of significant 
$\rho$ and would 
be due to dead stars, planets, neutron stars and other normal 
baryonic matter debris.
Each term within the series has $z$-dependence of $e^{-k_n|z|}$ which causes
the steep density fall-off profile as shown in
Figure \ref{fig:milkywaydensity}(b).
This fortifies the picture of a standard galactic disk-like shape as 
opposed to a halo sphere. From the evidence provided by rotation curves, there is no support for the widely accepted 
notion of massive halos of exotic dark matter surrounding visible 
galactic disks: conventional gravitational theory, namely general 
relativity, accounts for the observed flat galactic rotation curves 
linked to flattened disks with no exotic dark matter.

\begin{figure}
\begin{center}
\includegraphics[width=3in]{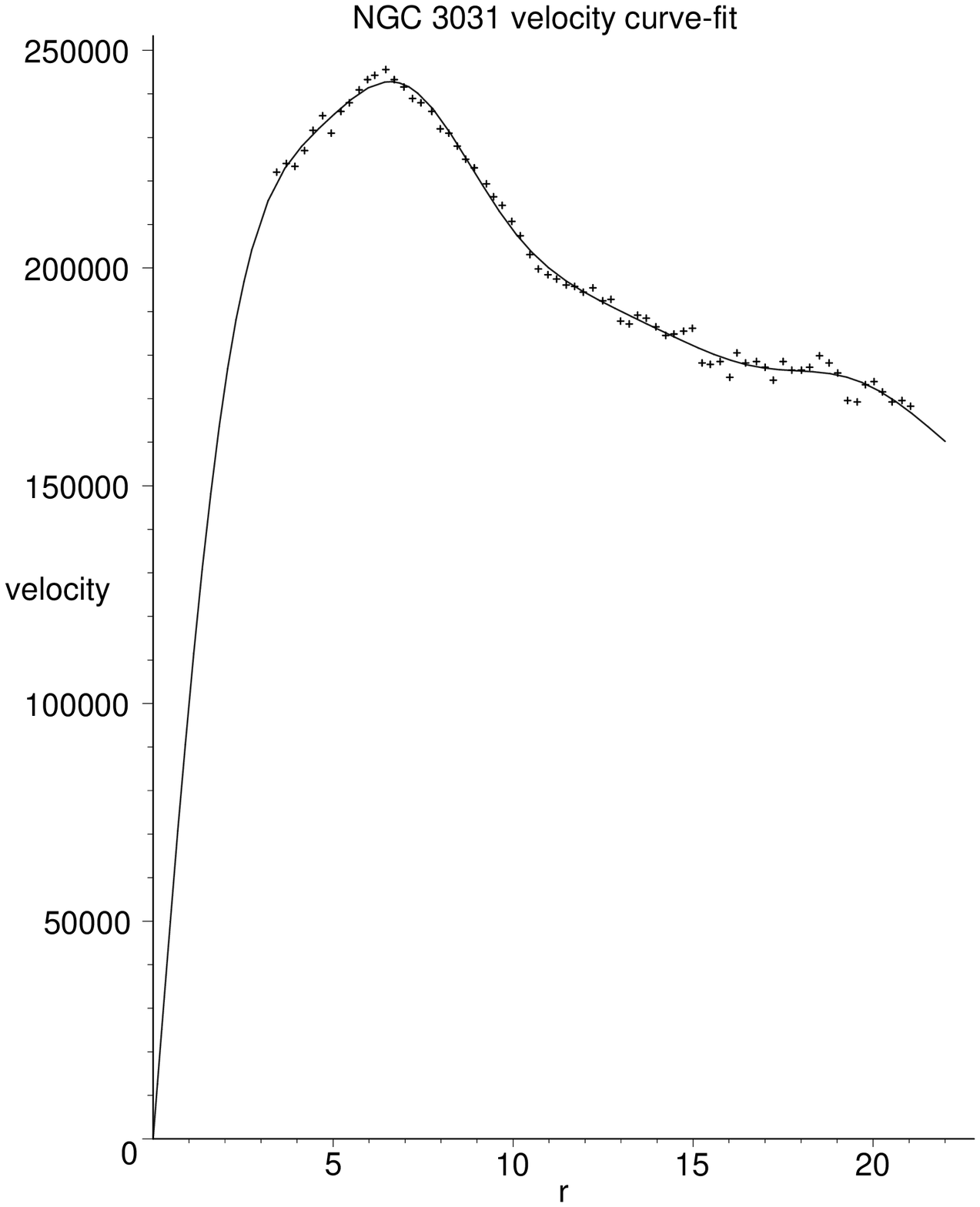}
\includegraphics[width=3in]{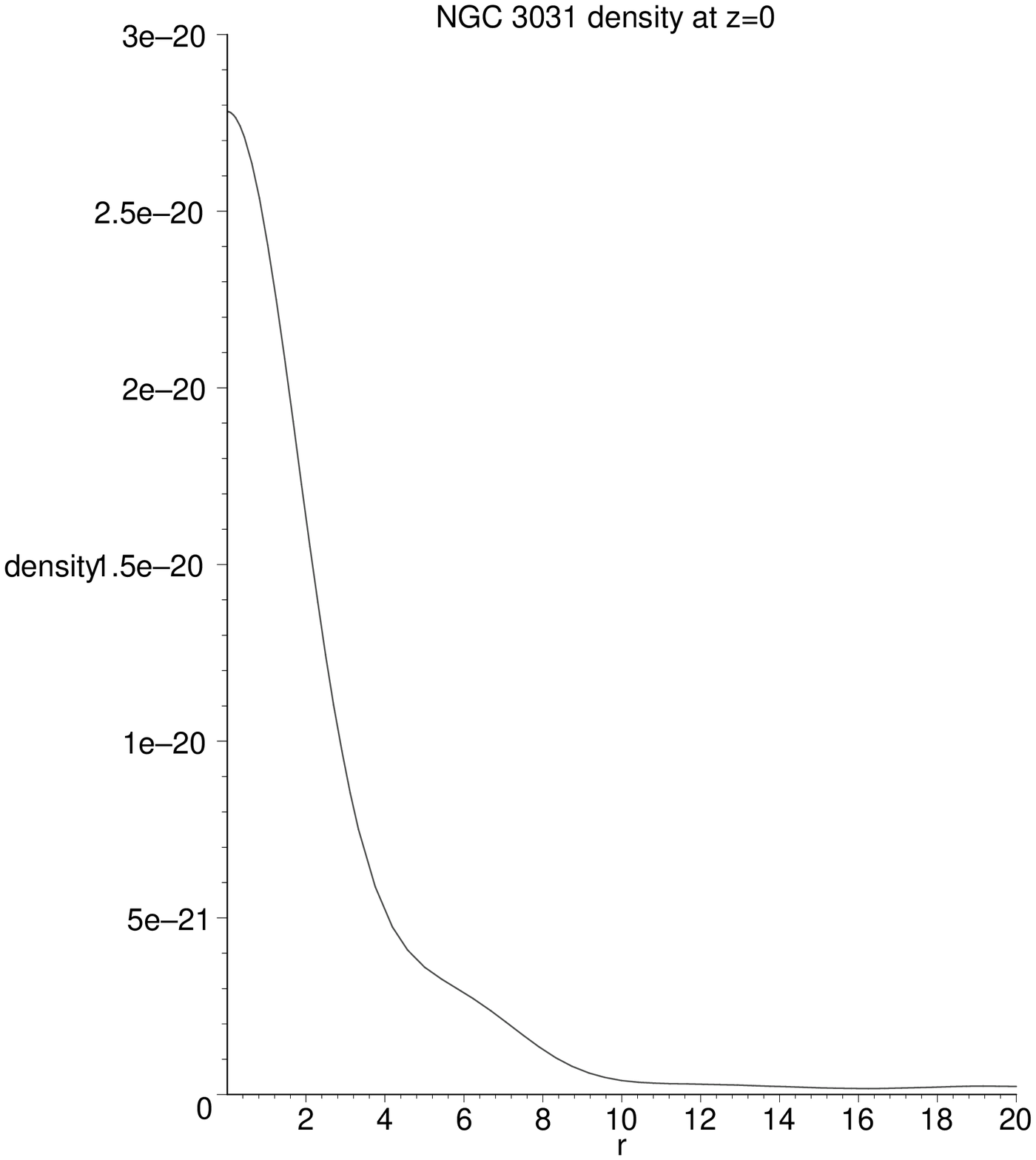}
\end{center}
\caption{
	Velocity curve-fit and 
	derived density for NGC 3031
}
\label{fig:ngc3031}
\end{figure}

\begin{figure}
\begin{center}
\includegraphics[width=3in]{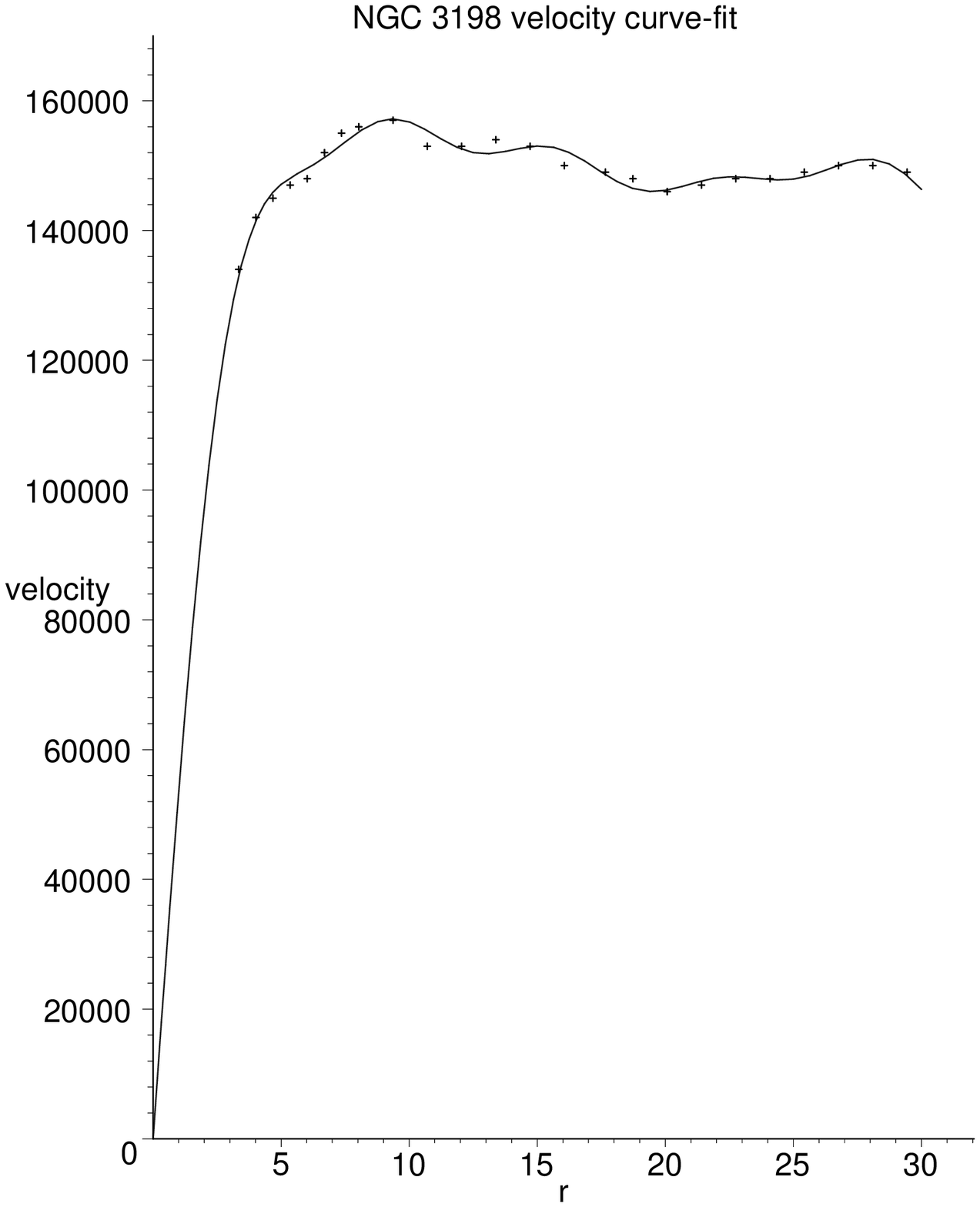}
\includegraphics[width=3in]{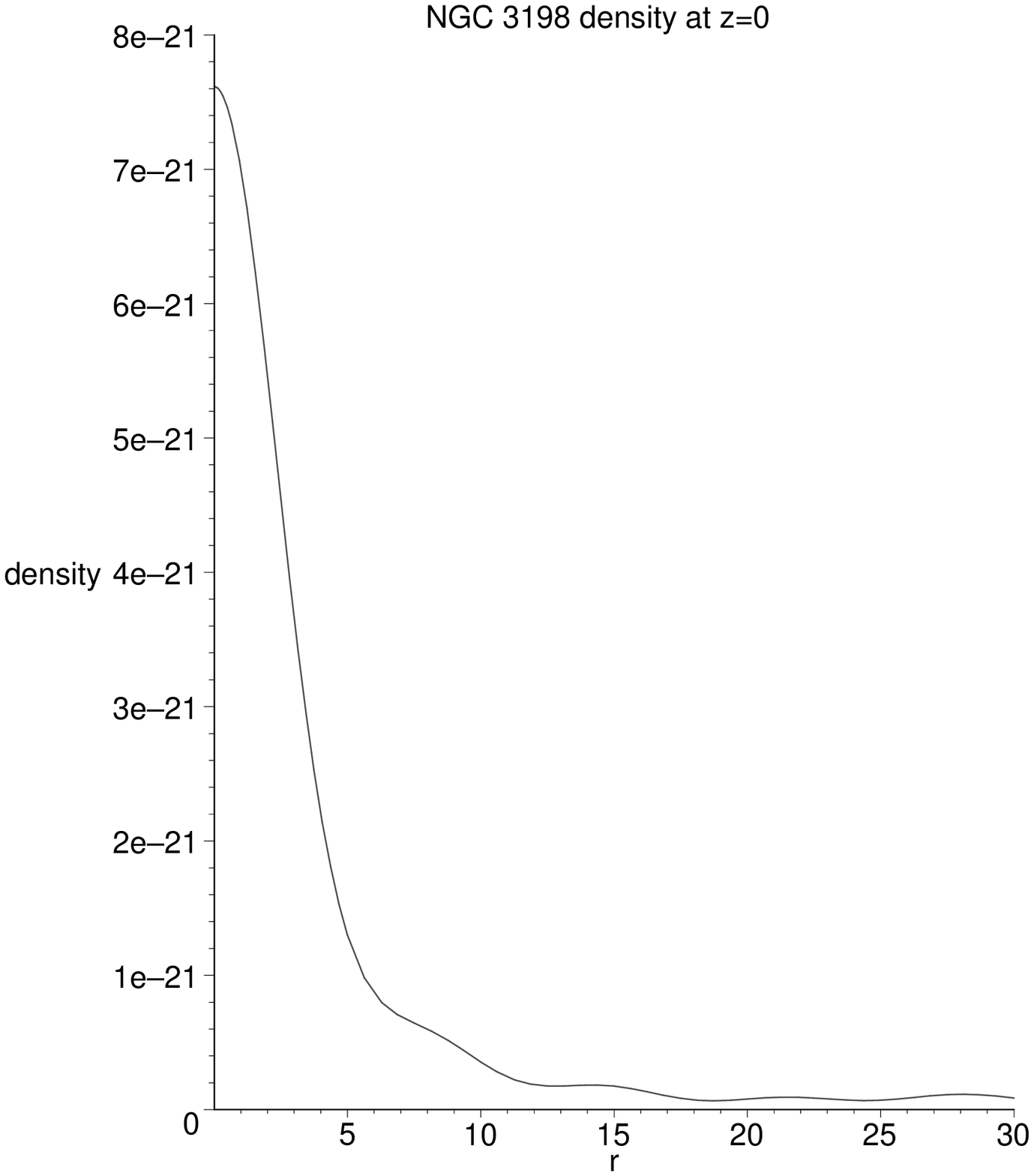}
\end{center}
\caption{
	Velocity curve-fit and 
	derived density for NGC 3198
}
\label{fig:ngc3198}
\end{figure}

\begin{figure}
\begin{center}
\includegraphics[width=3in]{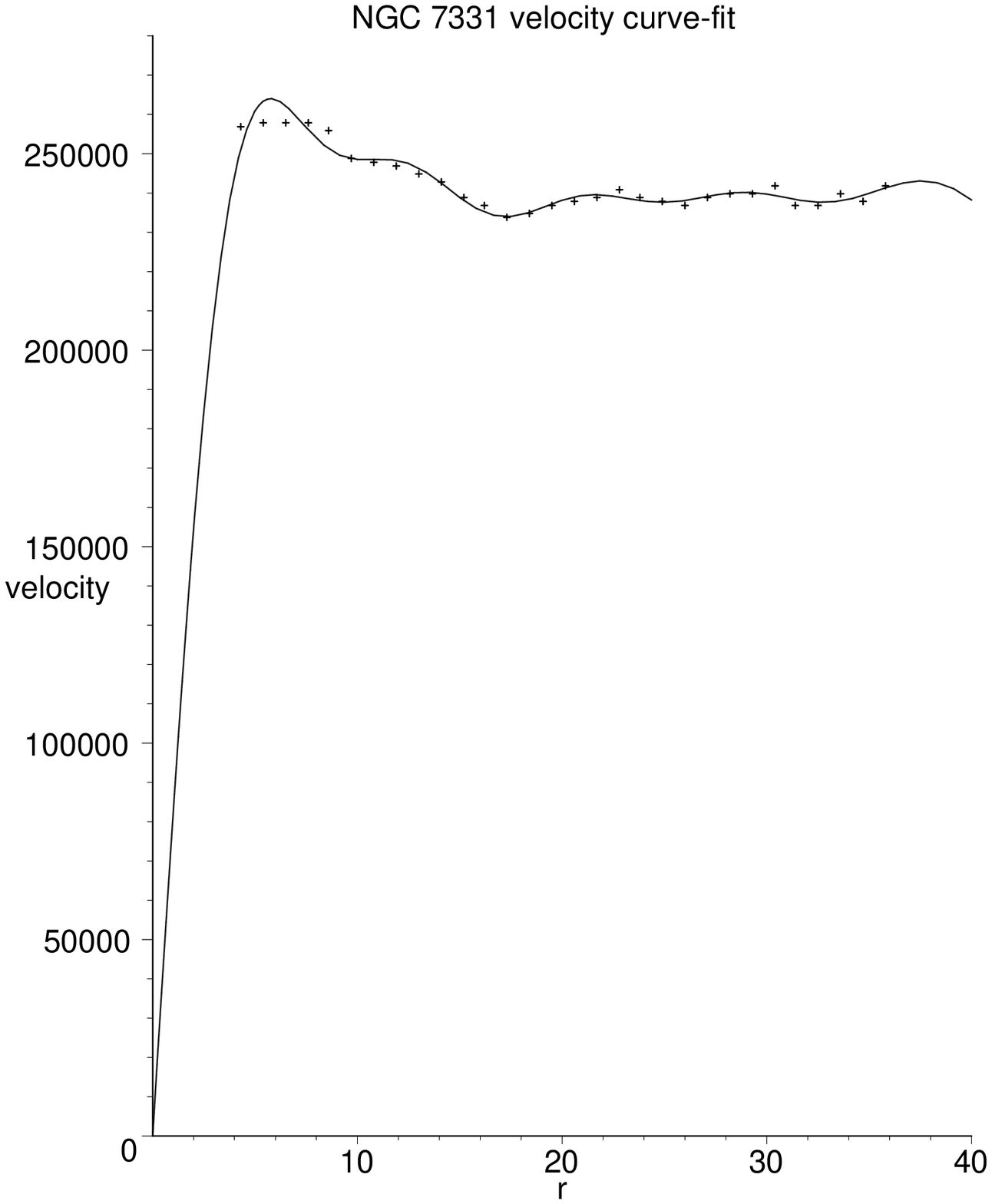}
\includegraphics[width=3in]{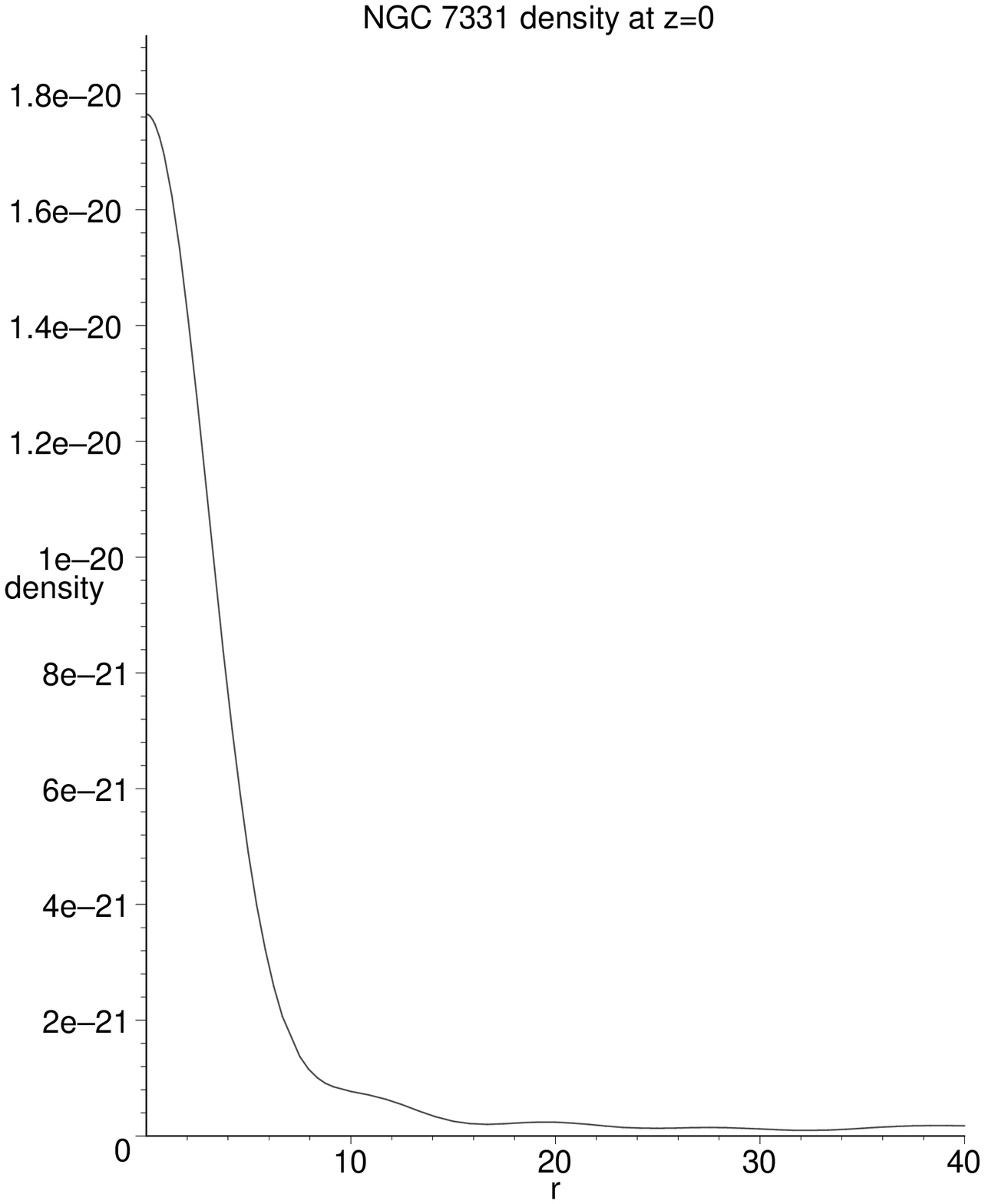}
\end{center}
\caption{
	Velocity curve-fit and 
	derived density for NGC 7331
}
\label{fig:ngc7331}
\end{figure}

We have also performed curve fits for the galaxies NGC 3031, NGC 3198 and
NGC 7331. The data are provided in the Appendix and the remarkably 
precise velocity curve fits are shown in Figures \ref{fig:ngc3031} 
to \ref{fig:ngc7331} where the density profiles are presented for 
$r$ at $z=0$. 
Again the picture is consistent with the observations and the mass
is found to be $10.1 \times 10^{10}M_\odot$ for NGC 3198. 
This can be compared to the result from Milgrom's \cite{bek} modified 
Newtonian dynamics of $4.9 \times 10^{10}M_\odot$ and the value 
given through observations (with Newtonian dynamics) by Kent 
\cite{Kent} of $15.1 \times 10^{10}M_\odot$.
While the visible light profile terminates at $r=14$ Kpc, the HI 
profile extends to 30 Kpc.  If the density is integrated to 14 Kpc, 
it yields a mass-to-light ratio of $7\Upsilon_\odot$. However, 
integrating through the HI outer region to $r=30$ Kpc 
yields $14\Upsilon_\odot$ using data from \cite{alba}. 

For NGC 7331, we calculate a mass of $26.0 \times 10^{10}M_\odot$. Kent 
\cite{Kent} finds a value of $43.3 \times 10^{10}M_\odot$. For NGC 3031, the mass is calculated to be $10.9\times 10^{10}M_\odot$ as compared to Kent's value of $13.3\times 10^{10}M_\odot$. Our masses are consistently lower than the masses projected by models invoking dark matter halos and our distributions roughly tend to follow the contours of the optical disks.

\begin{figure}
\begin{center}
\includegraphics[width=3in]{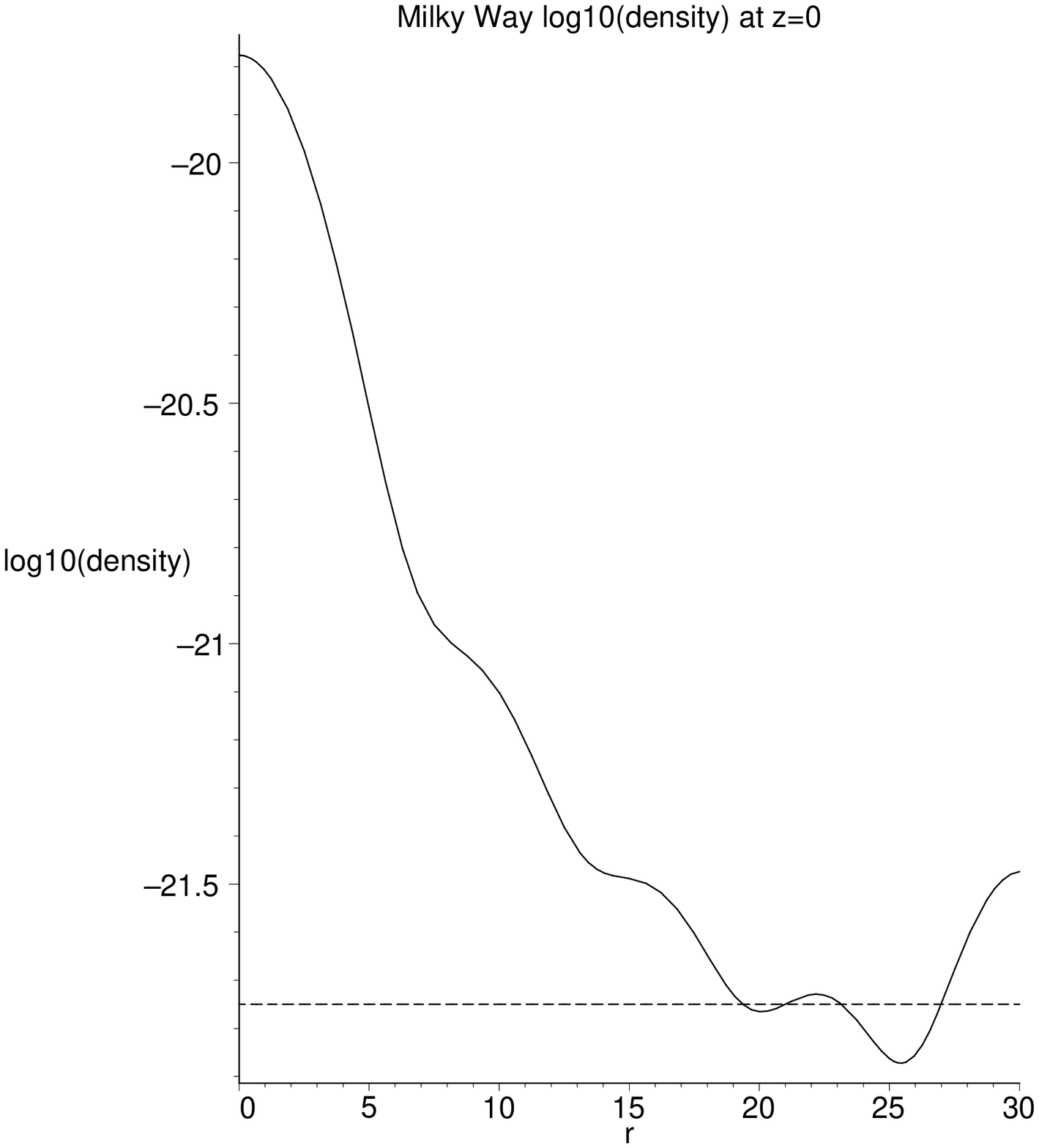}
\includegraphics[width=3in]{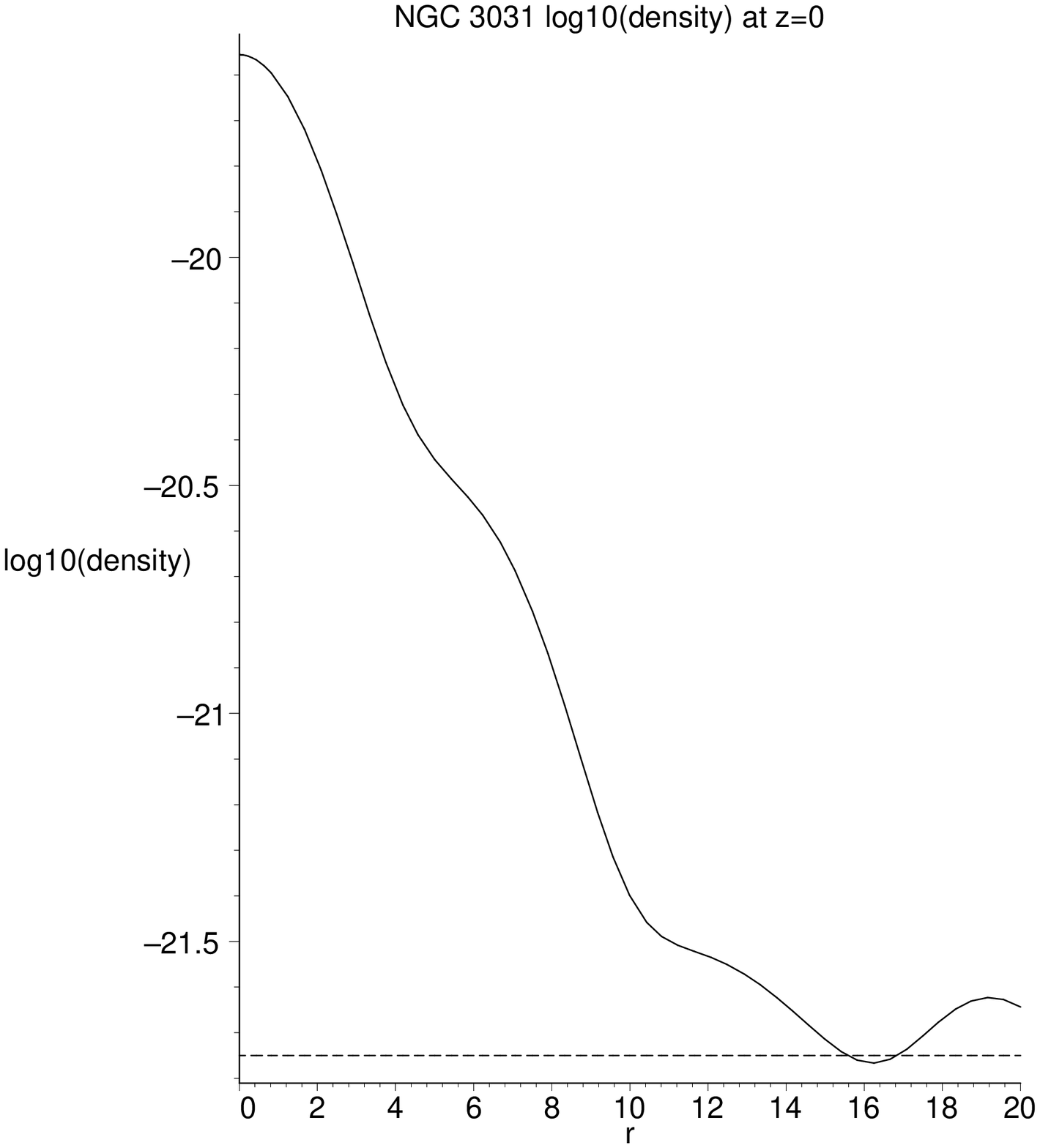}
\end{center}
\caption{
	Log graphs of density for the Milky Way (left) and
	NGC 3031 (right) showing the density fall-off.
	The $-21.75$ dashed line provides a tool to predict
	the outer limits of visible matter. The fluctuations at
	the end are the result of limited curve-fitting terms.
}
\label{fig:first2logdensity}
\end{figure}

\begin{figure}
\begin{center}
\includegraphics[width=3in]{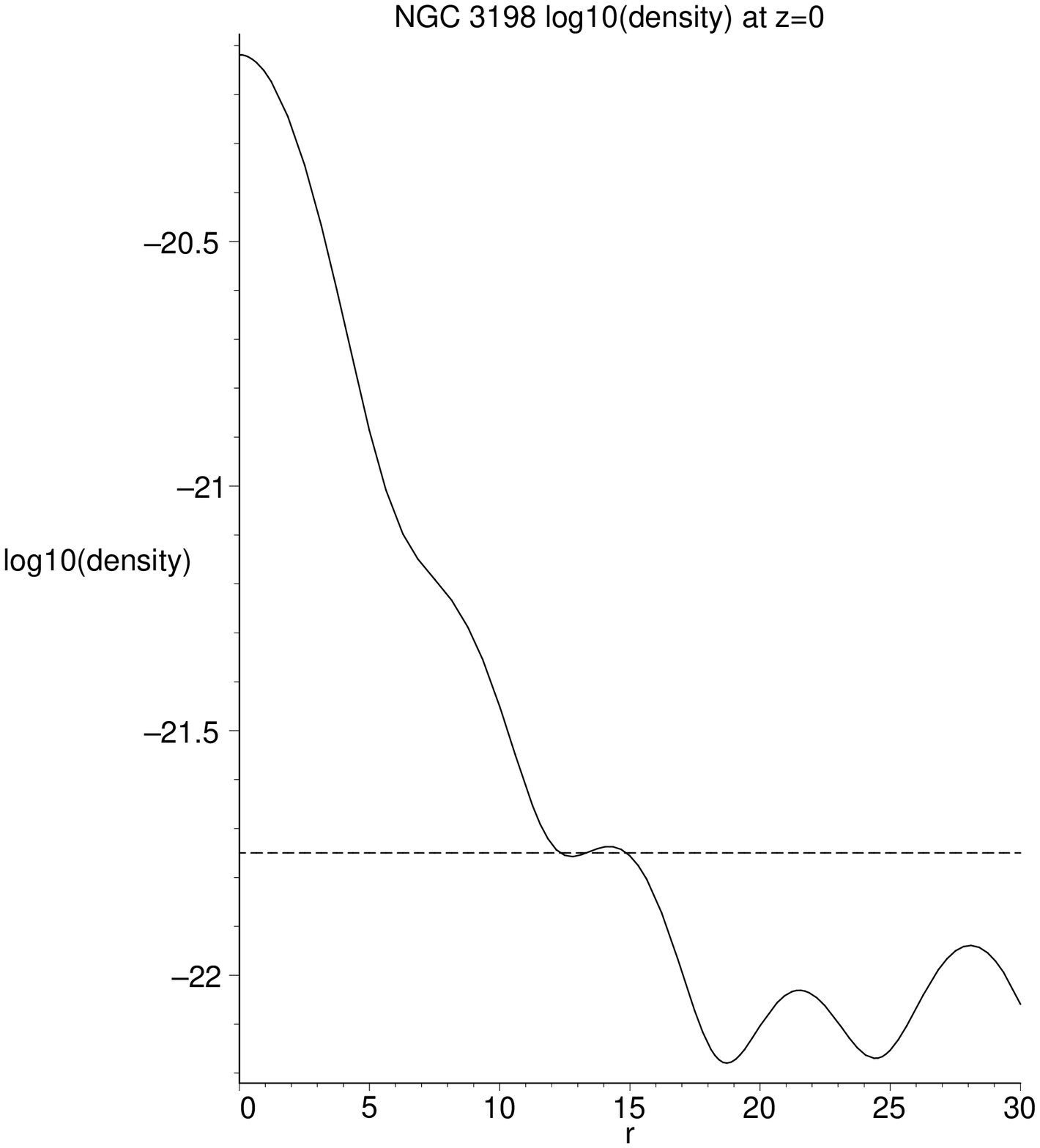}
\includegraphics[width=3in]{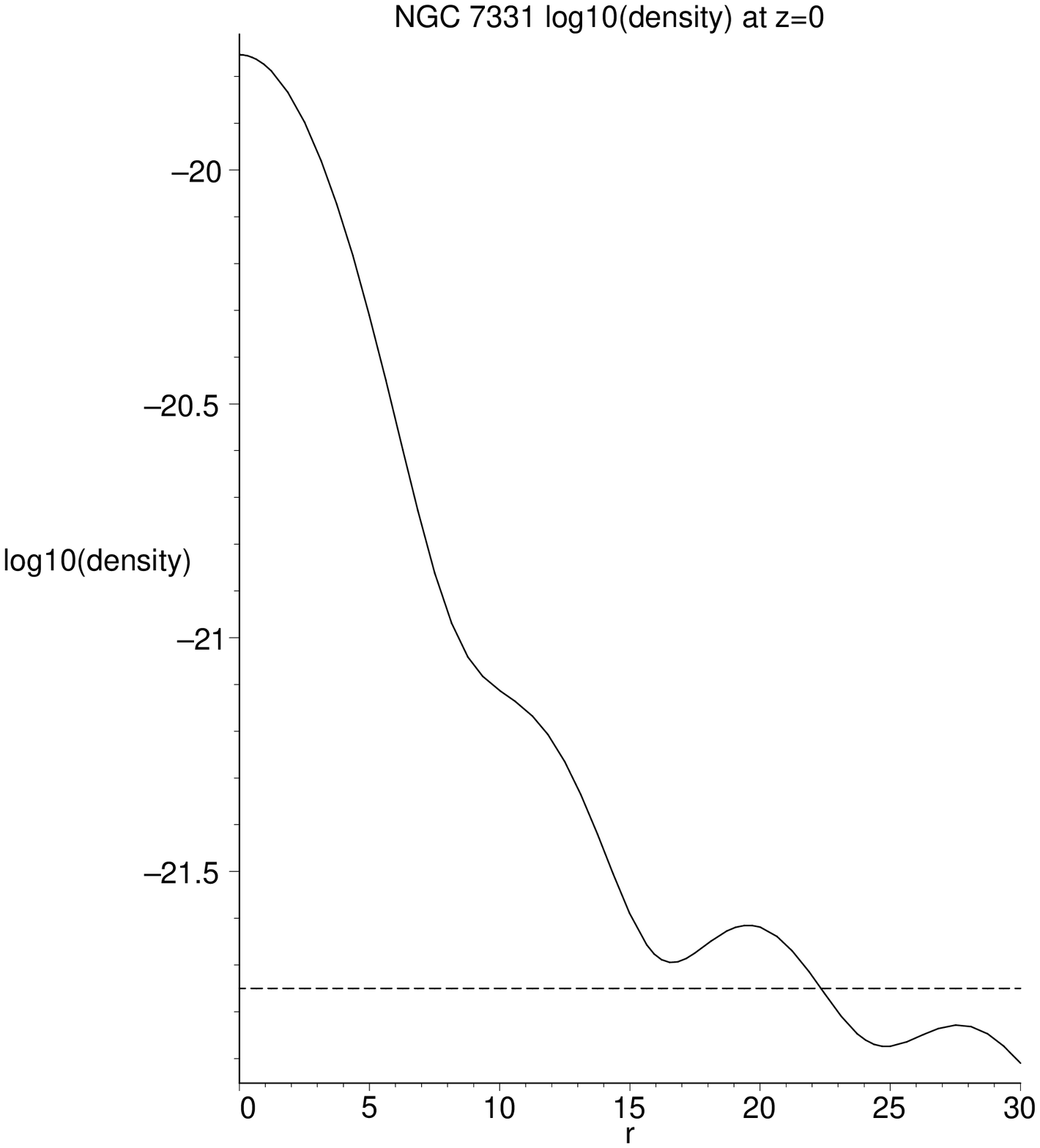}
\end{center}
\caption{
	Log graphs of density for the NGC 3198 (left) and
	NGC 7331 (right) showing the density fall-off.
	The $-21.75$ dashed line provides a tool to predict
	the limits of luminous matter. As before, there are
 fluctuations near the border.
}
\label{fig:last2logdensity}
\end{figure}

It is interesting to note that from the figures provided by Kent \cite{Kent}for optical intensity curves and our log density profiles for NGC 3031, NGC 3198 and NGC 7331, we determine that the threshold for the onset of visible galactic light is at  $10^{-21.75}$ kg$\cdot$m$^{-3}$ (Figure \ref{fig:first2logdensity} and Figure \ref{fig:last2logdensity}). It would be of interest to explore as many sources as possible to test the indicated hypothesis that this density is the universal optical luminosity threshold for galaxies. Alternatively, should this hypothesis be further substantiated, the radius at which the optical luminosity fall-off occurs can be predicted for other sources using this special density parameter. The predicted optical luminosity fall-off for the Milky Way is at a radius of 19-21 Kpc based upon the density threshold that we have determined.

Various authors attempt to incorporate the Tully- Fisher law \cite{tul} 
into their modified theories of gravity. General relativity can provide an 
equivalent albeit considerably more complicated relation but in integral form. From (\ref{Eq9a}) and 
(\ref{Eq11}), the radial gradient of the galactic mass can be expressed in 
terms of velocity as
\eqnn{Eq15}{
	M_r(r)
	=\frac{1}{2G}\int_{0}^{\infty}
	\left(r\left(V_r^2+V_z^2\right)
	+ \frac{V^2}{r} +2VV_r \right)\, dz
}
and a doubling has been used to account for the lower disk contribution. 

\section{Concluding Comments}

One might be inclined to question how this large departure from the 
Newtonian picture regarding galactic rotation curves could have 
arisen since the planetary motion problem is also a gravitationally 
bound system and the deviations there using general relativity are 
so small. The reason is that the two problems are very different: in 
the planetary problem, the source of gravity is the sun and the 
planets are treated as test particles in this field (apart
from contributing minor perturbations when necessary). They respond to 
the field of the sun but they do \textit{not} contribute to the 
field. By contrast, in the galaxy problem, the source of the field is 
the combined rotating mass of all of the freely-gravitating elements 
themselves that compose the galaxy. 

We have seen that the non-linearity for the computation of density 
inherent in the Einstein field equations for a stationary 
axially-symmetric pressure-free mass distribution, even in the case 
of weak fields, leads to the correct galactic velocity curves as 
opposed to the incorrect curves that had been derived on the basis of 
Newtonian gravitational theory. Indeed the results were consistent 
with the observations of velocity as a function of radius plotted as 
a rise followed by an essentially flat extended region and no halo of 
exotic dark matter was required to achieve them. The density 
distribution that is revealed thereby is one of a concentrated 
mass-density disk, in support of the ``maximum disk" ( see \cite{bos} 
and references therein) models but without an accompanying extended dark matter halo. With the ``dark" matter concentrated 
in the disk which is itself visible, it is natural  to regard the 
non-luminous material  as normal baryonic matter. 

It is unknown how far the galactic disks extend. More data points beyond those provided thus far by observational astronomers would enable us to extend the velocity curves further. Presumably a point (let us call it $r_f$) is reached where we can set $\rho$ to zero. At this point, (\ref{Eq2}) no longer applies as there are no longer co-rotating fluid elements being tracked. As a result, (\ref{Eq7}) no longer applies and the $w$ function is no longer constant. Beyond $r_f$, no further mass is accumulated. If we suppose that this is the case at the extremities of the HI regions indicated, then the masses that we have derived are indeed the total masses. \textit{It is to be emphasized that the flat rotation curves have been achieved with these modest mass values, without a massive exotic dark matter halo.}  

Nature is merciful 
in providing one linear equation that enables us by superposition to 
model disks of variable density distributions. This opens the way to 
studies of other sources and with further refinements.  
Moreover, it will be of interest to extend this general relativistic 
approach to the other relevant areas of astrophysics with the aim of 
determining whether there is any scope remaining for the presence of 
any exotic dark matter in the universe. Clearly the absence of such 
exotic dark matter would have considerable significance. 

\vskip 0.25in

{\small {\bf Acknowledgments:} This work was supported in part by a grant
from the Natural Sciences and Engineering Research Council of Canada.}

\clearpage
{\small 

\section{Appendix}

The coefficients for 
\eqn{
	N(r,z)= -\sum_{n=1}^{10} C_n k_n e^{-k_n |z|}J_1(k_nr)
}
are tabulated in Tables \ref{table:milkyway} to \ref{table:ngc7331}
with $r$ and $z$ in Kpc.  The velocity in m/sec is given by
\eqn{
	V(r,z)= \frac{3\times 10^8}{r} N(r,z)
}
and the density in kg/m$^3$ is given by
\eqn{
	\rho(r,z) = 5.64\times 10^{-14}
		\frac{ \left( N_r^2 + N_z^2 \right) }{r^2}
}

\begin{table}[ht]
\begin{center}
\begin{tabular}{c c}
\hline
$-C_n k_n$ & $k_n$ \\
\hline
 0.0012636497740 & 0.06870930165 \\
 0.0004520156256 & 0.15771651740 \\
 0.0001785404942 & 0.24724936890 \\
 0.0002946610499 & 0.33690098400 \\
 0.0000103378815 & 0.42659764880 \\
 0.0002127633340 & 0.51631611340 \\
-0.0000221015927 & 0.60604676080 \\
 0.0001346275993 & 0.69578490080 \\
-0.0000123824930 & 0.78552797510 \\
 0.0000666973093 & 0.87527447050 \\
\hline
\end{tabular}
\end{center}
\caption{Curve-fitted coefficients for the Milky Way}
\label{table:milkyway}
\end{table}



\begin{table}[ht]
\begin{center}
\begin{tabular}{c c}
\hline
$-C_n k_n$ & $k_n$ \\
\hline
0.0011694103480 & 0.1093102526 \\
0.0004356556836 & 0.2509126413 \\
0.0003677376760 & 0.3933512687 \\
0.0001484103801 & 0.5359788381 \\
0.0000837048346 & 0.6786780777 \\
0.0000414084713 & 0.8214119986 \\
0.0000429277032 & 0.9641653013 \\
0.0000550130755 & 1.1069305240 \\
0.0000238560073 & 1.2497035970 \\
0.0000129841761 & 1.3924821120 \\
\hline
\end{tabular}
\end{center}
\caption{Curve-fitted coefficients for NGC 3031}
\label{table:ngc3031}
\end{table}



\begin{table}[ht]
\begin{center}
\begin{tabular}{c c}
\hline
$-C_n k_n$ & $k_n$ \\
\hline
0.00093352334660 & 0.07515079869 \\
0.00020761839560 & 0.17250244090 \\
0.00022878035710 & 0.27042899730 \\
0.00009325578799 & 0.3684854512  \\
0.00007945062639 & 0.4665911784  \\
0.00006081834319 & 0.5647207491  \\
0.00003242780880 & 0.6628636447  \\
0.00003006457058 & 0.7610147353  \\
0.00001687931928 & 0.8591712228  \\
0.00003651365250 & 0.9573314522  \\
\hline
\end{tabular}
\end{center}
\caption{Curve-fitted coefficients for NGC 3198}
\label{table:ngc3198}
\end{table}



\begin{table}[ht]
\begin{center}
\begin{tabular}{c c}
\hline
$-C_n k_n$ & $k_n$ \\
\hline
0.0015071991080 & 0.0586542819 \\
0.0003090462519 & 0.1346360514 \\
0.0003960391396 & 0.2110665344 \\
0.0001912008955 & 0.2875984009 \\
0.0002161444650 & 0.3641687246 \\
0.0000988404542 & 0.4407576578 \\
0.0001046496277 & 0.5173569909 \\
0.0000619051218 & 0.5939627202 \\
0.0000647087250 & 0.6705726616 \\
0.0000457420923 & 0.7471855236 \\
\hline
\end{tabular}
\end{center}
\caption{Curve-fitted coefficients for NGC 7331}
\label{table:ngc7331}
\end{table}


\end{document}